\documentstyle[aps,epsf]{revtex}

\begin{document}

\title{Double-logarithmic asymptotics of the electromagnetic form
factors of the electron and quark}

\author{B.I. Ermolaev\\
 A.F. Ioffe
Physico-Technical Institute,
194021 St.Petersburg,  Russia \\
and\\
S.I. Troyan\\
St.Petersburg Institute of Nuclear Physics, 188350
St.Petersburg, Gatchina, Russia}

\maketitle

\begin{abstract}
The asymptotic behaviour of the electromagnetic form factors of the
electron and quark is obtained in the double-logarithmic approximation
for Sudakov kinematics, i.e. for the case that the value of the
momentum transfer is much greater than the mass of the particle.
\end{abstract}

\vspace{.5cm} 

\section{Introduction}

The interaction of an electron or quark with an
electromagnetic field is described in terms of two independent
form factors $f$ and $g$:

\begin{equation}
\Gamma_{\mu} = \bar{u}(p_2)\left[ \gamma_{\mu}f(q^2) - \frac{\sigma_{\mu
\nu}q_{\nu}}{2m}g(q^2)\right] u(p_1)
\label{1}
\end{equation}
where
$\sigma_{\mu \nu} = [\gamma_{\mu}\gamma_{\nu} - \gamma_{\nu}\gamma_{\mu}]/2$,
$q=p_2 - p_1$  is the momentum transferred to the electron or
the quark, $m$ is the electron or quark mass and
both  the form factors $f$ and $g$ depend on $q^2$.

In the Born approximation $f = 1$ and $g = 0$. The calculation of one-loop
radiative corrections \cite{s} shows that the form factor $f$ depends
on the infrared cutoff as well as on the ultraviolet one.
In contrast, the
form factor $g$ in the one-loop approximation is both ultraviolet-
and infrared-
stable. As the form factor $g$ contributes to  the
value of the anomalous magnetic moment, it has been calculated
with great accuracy by direct
graph-by-graph calculations in the case that $q^2 = 0$. The most recent review
of such results is given in \cite{cm}.
Meanwhile, the form factor $f$ was calculated many years ago  in
the ``opposite'' kinematical
region  of very large  momentum transfer:

\begin{equation}
-q^2 \gg m^2 ,
\label{2}
\end{equation}
in the leading logarithmic approximation (LLA), where the most important,
double logarithmic contributions to all orders in the QED coupling
$\alpha$ of the electron are taken into account.
The sum of such contributions, the double logarithmic (DL)
asymptotic behaviour of the form factor of the electron in the kinematical
region (\ref{2}), is

\begin{equation}
f =
\exp\left[-\frac{\alpha}{4\pi}\ln^2\left(\frac{-q^2}{m^2}\right)\right]
\label{3}
\end{equation}

The famous expression (\ref{3}), obtained by V.V. Sudakov \cite{sud},
was actually the birth of
the approach that is so popular at present -- the
double logarithmic approximation (DLA) -- where only the leading contributions
$\sim (\alpha \ln^2(-q^2))^n$ are taken into account to all orders of the
perturbation series.  The generalization of the Sudakov form factor of
Eq. (\ref{3}) to the quarks of QCD
obtained in \cite{sudqcd} amounts to replacing
$\alpha$ by $\alpha_s C_F$  in Eq. (\ref{3}),
where $C_F = (N^2-1)/2N = 4/3$ for the colour group $SU(3)$.

The exponential fall in Eq. (\ref{3}) as $-q^2$ increases corresponds to a
suppression of the non-radiative hard scattering of an electron by a virtual
photon.
The amplitude taking into account the
bremsstrahlung of $n$ ``soft'' photons
was shown in \cite{gorsh} to be a product of independent factors :

\begin{equation}
f_n =  B_1 B_2 ...B_n f(q^2) ,
\label{4}
\end{equation}
where the bremsstrahlung factors $B_i$  are given by (we drop the QED
coupling here)

\begin{equation}
B_ i =   \frac{p_2 l_i}{p_2k_i}  -\frac{p_1 l_i}{p_1 k_i}
\label{5}
\end{equation}
$l_i$ being the polarization vector and $k_i$ the momentum of the $i$
-th
emitted photon $(i = 1, ...,n)$. As $f$ in Eq.~(\ref{5}) does not
depend on $k_i$, and each of
$B_i$ does not depend on $k_j$ with $j \neq i$,  Eq. (\ref{5})
leads to a Poisson energy spectrum for bremsstrahlung photons
in the DLA.
The violation of the Poisson distribution in QCD for the emission of
``soft'' gluons in the DLA was obtained in \cite{fk} by calculating
Feynman graphs up to order $\alpha_s^2$.
The generalization of the form factor $f$ in Eq. (\ref{4}) to
QCD was given in \cite{ef}, \cite{efl}.

As follows from the lowest, one-loop approximation \cite{s}, the form
factor $g$ of the electron in the kinematic region (\ref{2}),

\begin{equation}
g^{(1)}(q^2) = -\frac{m^2}{q^2} \frac{\alpha}{\pi}
\ln\left(\frac{-q^2}{m^2}\right)~~,
\label{a}
\end{equation}
compared to the form factor $f$, lacks one large logarithmic factor and
is suppressed by the power factor $m^2/q^2$.  
However, in spin-flip scattering amplitudes 
the relatively large form factor $f$ is multiplied by the same small
factor $m^2/q^2$ and the effects of the form factor $g$ become
sizeable.

In the present work we calculate the form factor $g$ for the electron and
the quark in the
kinematical region (\ref{2}) in the double logarithmic approximation
(DLA).  Then, using the results of \cite{efl}, we obtain relations between
the radiative (inelastic) electromagnetic form factors of the
electron and quark.

The paper is organized as follows: in Section 2 we prove the
factorization property  for the case of the elastic form factor $g$
of the electron and calculate $g$ in DLA first by direct calculation
of Feynman graphs, then by constructing and solving infrared
evolution equations $f$ and $g$ obey.  In Section 3
we generalize
that result to QCD.  In Section 4 we show how the previous results
can be generalized to the case of the inelastic form factors of the
electron and quark. Section 5 is devoted to concluding remarks.

\section{Elastic form factors of the electron}

\begin{figure}
\begin{center}
\begin{picture}(140,200)
\put(0,10){
\epsfbox{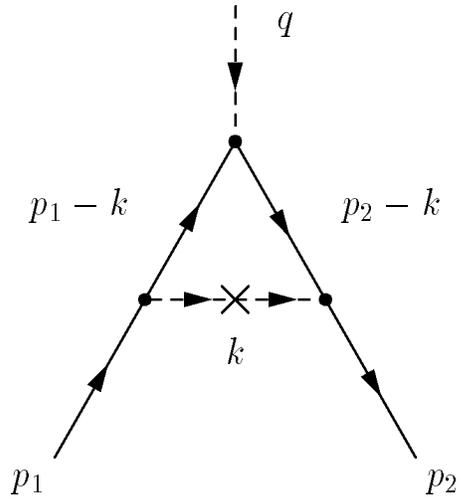}
}
\end{picture}
\end{center}
\caption{One-loop diagram for $\Gamma_{\mu}$.}
\label{oneloop}
\end{figure}

In the lowest, one-loop approximation $f$ and $g$ form factors  of electron
were calculated long ago in \cite{s}. In the Feynman gauge the only
graph yielding
the logarithmic contributions to $f$ and
$g$ in the kinematics (\ref{2}) is shown in Fig.~\ref{oneloop}.
The result is \cite{s}

\begin{equation}
g^{(1)}(q^2) = -\frac{m^2}{q^2} \frac{\alpha}{\pi} \ln 
\left(\frac{-q^2}{m^2}\right)~~.
\label{aa}
\end{equation}

In order to reproduce (\ref{a}) in the manner easy to generalize
to higher loops with the DL accuracy, we use the Sudakov
parameterisation \cite{sud}. According to it,
momentum $k_i$ of each soft virtual particle (a virtual photon)
is expressed through the longitudinal
variables
$\alpha_i$, $\beta_i$ and the transverse momentum $k_{i \perp}$:

\begin{equation}
k_i = \alpha_ip'_2 + \beta_i p'_1 + k_{i \perp}
\label{sudab}
\end{equation}

where $(k_{i \perp}p_1) = (k_{i \perp}p_2) = 0$, $s\equiv 2(p_1p_2)
\approx -q^2$ and
$p_1$, $p_2$ are the momenta of the electron (see Fig.~\ref{oneloop}). Then,

\begin{equation}
p'_1 = p_1 - (m^2/s)p_2, \qquad p'_2 = p_2 - (m^2/s)p_1
\label{defq}
\end{equation}
so that $(p'_1)^2 = (p'_2)^2 = 0$.

Applying the Feynman rules to the graph in Fig.~\ref{oneloop}, we obtain the
expression

\begin{eqnarray}
\Gamma_1 = -\imath (\alpha/4\pi^3) \int dk_1
\bar{u}(p_2)\gamma_{\nu}\frac{(\hat{p}_2 + m - \hat{k}_1)}
{[(p_2 - k_1)^2 - m^2 + \imath\epsilon]} \nonumber \\
\gamma_{\mu}
\frac{(\hat{p}_1 + m - \hat{k}_1)}
{[(p_1 - k_1)^2 - m^2+ \imath\epsilon]}
 \gamma_{\nu} u(p_1)~
\frac{1}{[k^2 + \imath\epsilon]} .
\label{m1}
\end{eqnarray}

Simplifying the spinor structure in Eq.~(\ref{m1}) and
integrating it over $k_{1 \perp}$ with the replacement
\footnote{This   replacement is correct for calculations in
double-logarithmic and single-logarithmic approximations.
One can prove it calculating the integral of Eq.~(\ref{m1}) by taking
residues.}

\begin{equation}
\frac{1}{k_1^2 + \imath\epsilon} \rightarrow
-2\imath\pi\delta(s\alpha \beta - k_{1\perp}^2)
\label{soh}
\end{equation}
so that

\begin{equation}
\int dk_1 \frac{F(k_1)}{k_1^2 + \imath\epsilon} \approx
 -2\imath\pi\int_0^1 d\alpha_1 \int_0^1d\beta_1 F(\alpha_1,~ \beta_1)
\Theta(s\alpha_1\beta_1 - \mu) ,
\label{sohsud}
\end{equation}

with $\mu$ being an infrared cutoff defined as
\begin{equation}
\label{mu}
k_{1\perp} > \mu \geq m,
\end{equation}
we arrive at

\begin{eqnarray}
\Gamma_1 = && -\left(\gamma_{\mu} \right) \frac{\alpha}{2\pi}
\int \frac{d\alpha_1}{(\alpha_1 - (m^2/s) \beta_1)}
\frac{d\beta_1}{(\beta_1 - (m^2/s) \alpha_1)}
\Theta(s\alpha_1 \beta_1 - \mu^2) + \nonumber \\
&& \frac{\alpha}{4\pi}
2m \left(\sigma_{\mu \nu} q_{\nu} \right)
\int \frac{d\alpha_1}{(\alpha_1  - (m^2/s) \beta_1)}
\frac{ d\beta_1}{(\beta_1 - (m^2/s) \alpha_1)}
[\alpha_1 (1 - \alpha_1) + \beta_1 (1 - \beta_1)]
\Theta(s\alpha_1 \beta_1 - \mu^2)~~.
\label{m1sud}
\end{eqnarray}

We have used in Eq.(\ref{m1sud}) the
notations
$\left(\gamma_{\mu}\right) \equiv \bar{u}(p_2)\gamma_{\mu}u(p_1)$ ,
$ \left( \sigma_{\mu \nu} q_{\nu} \right) \equiv (1/2) \bar{u}(p_2)
\sigma_{\mu \nu} q_{\nu} u(p_1)$.

Let us comment on  Eq.(\ref{m1sud}).
We retained only leading contributions to the form factors $f$ and $g$
in Eq.(\ref{m1sud}). The leading contribution to $f$ is the
double logarithmic
one, with one logarithm as a result of integrating over $\alpha_1$ and
the other logarithm over $\beta_1$. As the denominator in Eq.(\ref{m1sud}) 
is, in essence,
just $\alpha_1\beta_1$, obtaining the two logarithms in
the leading contribution to $f$
corresponds to dropping all the Sudakov variables in the numerator
of  Eq.(\ref{m1sud}).
In order to get the leading contribution to $g$ one
must not neglect $\alpha_1$, $\beta_1$ in the numerator. Keeping either
$\alpha_1$ or $\beta_1$ in the numerators of  Eq. (\ref{m1sud})
immediately kills one logarithmic contribution and keeping products 
$\alpha_1\beta_1$ kills the both logarithms. In the present paper we
discuss only the leading contribution to $g$, therefore through the
paper we neglect contributions containing
products of $\alpha_i$ and $\beta_i$ (i = 1,2,..). In particular,
it means that we neglect contributions containing powers of
$k_{\perp}$ because
$k^2_{1~\perp} = s\alpha_1\beta_1$. However, we account for the terms in the
numerator of  Eq.(\ref{m1sud}) containing powers of $\alpha_1$
and of $\beta_1$ separately.
Then, as $m$ is the small parameter in the kinematics (\ref{2}),
we neglected in the numerators of Eq.(\ref{m1sud}) contributions
of order $m$ to the
form factor $f$  and contributions of order $m^2$ to $g$ .
We will do the same neglections when we consider the higher loop
contributions to those form factors.\\
Comparing Eq.~(\ref{m1sud}) and (\ref{1}), we see that
the contribution of $\Gamma_1$ to the form factor $f$ is

\begin{equation}
-\frac{\alpha}{2\pi} \left(\gamma_{\mu}\right)
\int \frac{d\alpha_1}{[\alpha_1 - (m^2/s) \beta_1]}
\frac{d\beta_1}{[\beta - (m^2/s) \alpha_1]} ~.
\label{f1sud}
\end{equation}

The DL region of integration in
Eqs.~(\ref{m1sud}) and (\ref{f1sud}) is

\begin{equation}
1 >\alpha_1 > (m^2/s),~~ 1> \beta_1 > (m^2/s),~~
\alpha_1\beta_1 > \mu^2/s .
\label{region}
\end{equation}

Obviously, the integral in Eq.~(\ref{f1sud}) diverges when
 $\mu = 0$,
so the  cut-off $\mu$ has to be kept .
As within DL accuracy one can choose an arbitrary infrared cut-off, we
choose, for the sake of simplicity,

\begin{equation}
\mu = m~.
\label{cutoff}
\end{equation}

After that we immediately reproduce the well known
one-loop result $f^{(1)}$
for the form factor $f$ with the DL accuracy:

\begin{equation}
f^{(1)} = - (\alpha/4\pi) \ln^2(-q^2/m^2)
\label{f1}
\end{equation}

In the contrast to $f$-form factor, the contribution to
form factor $g$ given by
the second integral in Eq.~(\ref{m1sud}), is infrared  stable.
Indeed, using the symmetry in
$\alpha_1, \beta_1$ we rewrite the second
integral in  Eq.~(\ref{m1sud}) with the
logarithmic accuracy as

\begin{eqnarray}
\int_0^1 \frac{d\beta_1}{\beta_1} 2 \beta_1(1 - \beta_1)
\int_{m^2/s}^1 \frac{d\alpha_1}{\alpha_1} =
\ln(s/m^2)~.
\label{integral}
\end{eqnarray}

Combining the result with the remaining numerical
factors in Eq.~(\ref{m1sud}) and
using the definition (\ref{1}) of form factor $g$ we reproduce
Eq.~(\ref{a}).

\begin{figure}
\begin{center}
\begin{picture}(180,200)
\put(0,10){
\epsfbox{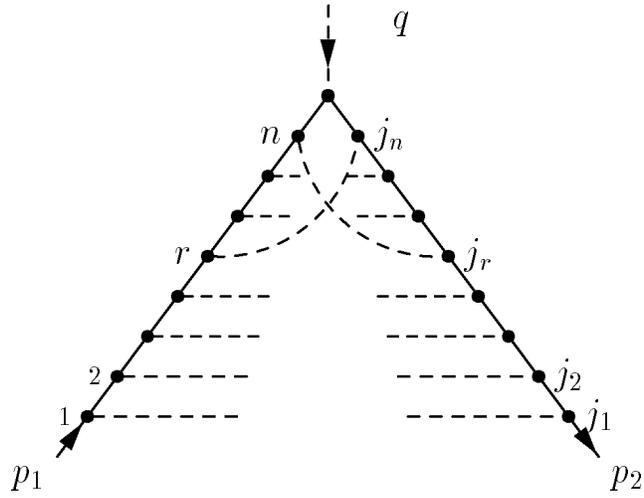}
}
\end{picture}
\end{center}
\caption{General diagram for $\Gamma_{\mu}$ in DLA.}
\label{mloop}
\end{figure}

Now let us consider the DL radiative corrections to the one-loop results
(\ref{f1}), (\ref{a}) for  $f$ and $g$ coming from
an arbitrary $\alpha^n$- order graph. In the Feynman gauge that
we use through
the present paper, graphs yielding the DL contributions
are  obtained from the one-loop graph in Fig.~\ref{oneloop}
by adding to it the photon propagators connecting
the lines with momenta $p_1$ and $p_2$ in an arbitrary way. However, such
graphs can not have the closed fermion loops because the loops yield the
single-logarithmic contribution only.
We find it convenient to regard the virtual photon propagators
connecting the electron lines with momenta $p_1$ and $p_2$
as the ones beginning on the line with $p_1$ and ending on the line with
$p_2$. Without loss of generality, we can numerate
with numbers $1,2,...,n$
the virtual photons
emitted from $p_1$-line so that the propagator
of the photon
with momentum $k_1$ begins at the bottom (see Fig.~\ref{mloop}). However, basically
those emitted photons do  not arrive at the $p_2$- line in the same order.
For an arbitrary
graph, the order for the $p_2$- line is $ j_1, j_2,...,j_n$
(see Fig.~\ref{mloop}). Thus, any of the graphs contributing to $\Gamma_{\mu}$
correspond to a certain permutation $j_n,...,j_1$ of the initial order
$n,...,1$.
In particular, the permutation $j_n,...,j_1$
with $ j_1 =1, j_2 = 2,.., j_n = n$ corresponds the ladder graph
and the other, non-ladder
graphs are obtained from it by  permutations of  numbers $(n,n-1,...,1)$ on
the $p_2$-line. So, in order $\alpha^n$ there are $n!$ different graphs
yielding the DL contributions. These graphs yield equal contributions
to the form factor $f$ when the electron is off-shell but for the
on-shell electron it does not hold.

Again without loss of generality, we
assume that the photon with number $j_n$ on the $p_2$-line has
number $r$ on line $p_1$ and the photon
with number $n$ on line $p_1$
has number $j_r$ on line $p_2$ as it shown in Fig.~\ref{mloop}.
It leads to the matching

\begin{equation}
k_{j_n} = k_r,~~k_{j_r} = k_n
\label{biggest}
\end{equation}

We do not need to do matchings between other emitted and absorbed photons,
with non-largest Sudakov variables, so we
let them be arbitrary. We demonstrate it in Fig.~\ref{mloop} by leaving disconnected
 those
photon lines.

Straightforward
applying the Feynman rules
to the graph in Fig.~\ref{mloop} yields

\begin{eqnarray}
\Gamma_n =&& (- \imath)^n \frac{\alpha^n}{(4\pi)^{3n}}
\int dk_1...dk_n \bar{u}_2
\gamma_{\lambda_{j_1}}
\frac{(\hat{p}_2 + m - \hat{k}_{j_1})}{[(p_2 - k_{j_n})^2 - m^2]}...
\gamma_{\lambda_{j_1}} \nonumber \\ &&
\frac{(\hat{p}_2 + m -\hat{k}_{j_n} -...- \hat{k}_{j_1})}
{[(p_2 - k_{j_1} - ... - k_{j_n})^2 - m^2]}
 \gamma_{\mu}
\frac{(\hat{p}_1 + m - \hat{k}_1 -...- \hat{k}_n)}
{[(p_1 - k_1 - ... -k_n)^2 - m^2]} \gamma_{\lambda_n} \nonumber \\ &&
\dots\frac{(\hat{p}_1 + m - \hat{k}-n)}{[(p_1 - k_1)^2 - m^2]}
\gamma_{\lambda_1}
 u_1 \prod_{j=0}^{j=n}\frac{1}{k^2_j}   ~.
\label{mn}
\end{eqnarray}

In terms of the Sudakov variables

\begin{equation}
\prod dk_i = (s/2)^n
\prod d\alpha_i d\beta_i d^2k_{i\perp}.
\label{dk}
\end{equation}

In (\ref{dk}) $i$ also runs from 1 to $n$.
Integration over every transverse momentum
$k_{i\perp}$ can be done with replacement
$1/[k^2+\imath\epsilon]$ by $-\imath2\pi\delta(k^2_i)$ just like it was done
for the one-loop integration (\ref{m1sud}). After that the integrand in
Eq.~(\ref{mn}) depends on $\alpha_i$, $\beta_i$.
In the DLA we must keep all the Sudakov variables
$\alpha_j$, $\beta_j$, $(j = 1,...,n)$
 in the denominators of
Eq.~(\ref{mn}). It is possible only if they obey to the following
ordering:

\begin{eqnarray}
\alpha_1<\alpha_2<... < \alpha_{n-1} < \alpha_1 < 1~, \nonumber \\
\beta_{j_1} < \beta_{j_2} <... < \beta_{j_n} < 1~.
\label{order}
\end{eqnarray}

Eq. (\ref{order}) reads that all $\alpha_i$,$\beta_i$ are small and
therefore we can drop all $k_i$ in the numerators in (\ref{mn})
compared to $p_1$, $p_2$ and at the same time keep only linear in
$\alpha_i, \beta_j$ contributions in the denominator.
When we do it, we arrive at the double- logarithmic
$\alpha^n$-th order contribution to the form factor $f$ :

\begin{equation}
f^{(n)} = (-1)^n \left(\frac{\alpha}{2\pi}\right)^n \int\prod
\frac{d\alpha_i d\beta_i}{\alpha_i \beta_i}
\Theta(s\alpha_i\beta_i - \mu^2)~.
\label{intf}
\end{equation}

However, contributions to $g$ come from the terms
in the numerator of Eq. (\ref{mn})
containing the Sudakov variables. Due to the ordering (\ref{order}),
the leading contributions of the numerator of (\ref{mn}) to $g$
come from the biggest of them,
depending on  $\alpha_n$ and on $\beta_{j_n}$.
So, we keep only such contributions and neglect the others.
For the self-consistency of the approach,
we can approximate the denominators in (\ref{mn}) by the
contributions linear in $\alpha_i$, $\beta_i$ for $f$
but we must keep all the powers of $\alpha_n$ and $\beta_{j_n}$
for contributions to $g$. At the same moment we can use linear
approximation for other $\alpha_i$, $\beta_i$ with $i\neq n,~j_n$.
So, in order to get the leading
contribution to $g$,  we approximate the numerator in (\ref{mn})
(cf. (\ref{m1sud})) by

\begin{equation}
(- 2m) \left(\sigma_{\mu \nu}q_{\mu}\right)
(2s)^{(n-1)}
\left[\beta_{j_n}(1 - \beta_{j_n})^{(n +1 - r)} +
\alpha_n(1 - \alpha_n)^{(n + 1 - j_n)}\right] ,
\label{nom}
\end{equation}

and the denominator, correspondingly, by

\begin{equation}
\alpha_n(1 - \alpha_n)^{(n-j_n)}\beta_{j_n}(1 - \beta_{j_n})^{(n- r)}
{\prod}^{\prime} \alpha_i\beta_i.
\label{denomg}
\end{equation}

We have used the notation ${\prod}^{\prime}$
 to show that this product does not include $\beta_{j_n}$ and
$\alpha_n$ .  We have neglected in Eq.~(\ref{nom}), just like we did
for Eq.(\ref{m1sud}), contributions containing products of $\alpha_n$
and $\beta_{j_n}$ as well as
terms containing $k_{i \perp}$, (i = 1,2,..,n) because accounting
for them leads to loss more than one logarithmic contribution.
Combining Eqs.~(\ref{mn}),(\ref{dk}),(\ref{denomg}) and
(\ref{nom}), we obtain $\alpha^n$ -order contributions of the graph
in Fig.~\ref{mloop} to $g$ ~:

\begin{eqnarray}
g^{(n)} = && -(-1)^n \left(\frac{\alpha}{2\pi}\right)^n 2\int d
 \beta_{j_n} d\alpha_n \frac{[\beta_{j_n}(1 - \beta_{j_n}) + \alpha_n(1
- \alpha_n)]} {\beta_{j_n} \alpha_n} \int \frac{d\beta_n
d\alpha_r}{\beta_n \alpha_r} \Theta(s\alpha_r\beta_r - \mu^2) \nonumber
\\       && \Theta(s\alpha_n\beta_n - \mu^2)
{\prod}^{\prime}\int\frac{d\alpha_id\beta_i}{\alpha_i\beta_i}
\Theta(s\alpha_i\beta_i - \mu^2)
\label{intg}
\end{eqnarray}

Using the symmetry of Eq.(\ref{intg}) in replacing
$\alpha_n \rightleftharpoons \beta_{j_n}$,
$\alpha_n \rightleftharpoons \beta_{j_n}$, we rewrite Eq. (\ref{intg}) as

\begin{eqnarray}
g^{(n)} = && -(-1)^n \left(\frac{\alpha}{2\pi}\right)^n
2\int d \beta_{j_n} 2(1 - \beta_{j_n})
\int \frac{d\alpha_{j_n}}{\alpha_{j_n}}
\Theta(s\alpha_{j_n}\beta_{j_n} - \mu^2) \nonumber \\ &&
\int \frac{d\beta_n d\alpha_r}{\beta_n \alpha_r}
{\prod}^{\prime}\int\frac{d\alpha_id\beta_i}{\alpha_i\beta_i}
\Theta(s\alpha_i\beta_i - \mu^2)
\label{master}
\end{eqnarray}

The integration regions in Eqs.(\ref{master}),(\ref{intf}) are identical.
They are defined by the ordering
(\ref{order}) and by the $\Theta$-functions. Thus, the only difference
between the integrals in  Eqs.(\ref{master}),(\ref{intf}) is the
integration over $\beta_r$ . The integration region over $\alpha_i$,
 $(i = 1,..,n)$ is

\begin{equation}
\frac{\mu^2}{s} \beta_i < \alpha_i < \alpha_{i + 1}
\label{lima}
\end{equation}

where the upper limit for intergrating over $\alpha_n$ is 1. The region
(\ref{lima}) is the same for any graph contributing to $f$, $g$.
Integration over $\beta_{j_l}$ runs from $\mu/^2/s$ to $\beta_{j_{l + 1}}$ ,
with $\beta_{j_{n+1}} = 1$. It leads to

\begin{equation}
f^{(n)} = (-1)^n \left(\frac{\alpha}{2\pi}\right)^n
\int_{\lambda}^1 \frac{d\beta_{j_n}}{\beta_{j_n}}
F^{(n)}(\beta_{j_n}, \lambda)
\label{mastf}
\end{equation}

where
\begin{equation}
F^{(n)}(\beta_{j_n}, \lambda) =  F^{(n)}(\beta_{j_n}/ \lambda) =
\int_{\lambda}^1 \frac{d\beta_{j_{n-1}}}{\beta_{j_{n-1}}}...
\int_{\lambda}^1 \frac{d\beta_{j_1}}{\beta_{j_1}}
\int_{\lambda/\beta_n}^{1} \frac{d\alpha_n}{\alpha_n}...
\int_{\lambda/ \beta_1}^{\alpha_2} \frac{d\alpha_1}{\alpha_1}
\label{F}
\end{equation}
and $\lambda = \mu^2/s$.

We have arranged the order of integrations in (\ref{mastf}),(\ref{F})
so that Eq.~(\ref{F}) looked similar
to Eq.~(\ref{intg}) for the contribution to
$g^{(n)}$ where only the  last integration over $\beta_{j_n}$ does
not yield a logarithmic contribution whereas each integration in
$F$ yields it (see(\ref{F})).

Now, using (\ref{F}), we can express $g^{(n)}$ given by
 Eq. (\ref{master})
in terms of $F$ :

\begin{equation}
g^{(n)} = -(-1)^n 2\left(\frac{\alpha}{2\pi}\right)^n
\int_{\lambda}^1 d\beta_{j_n}2(1 - \beta_{j_n})
F^{(n)}(\beta_{j_n}/\lambda) .
\label{mastg}
\end{equation}

Replacing $\beta_{j_n}$ in Eqs.~$(\ref{mastf}),(\ref{mastg})$,
by $\beta$, we rewrite  them as

\begin{equation}
f^{(n)} = (-1)^n \left(\frac{\alpha}{2\pi}\right)^n U^{(n)}_f
\label{fn}
\end{equation}
where

\begin{equation}
U^{(n)}_f(\lambda) = \int_{\lambda}^1 \frac{d\beta}{\beta}
F^{(n)}(\beta/\lambda) ,
\label{uf}
\end{equation}
and

\begin{equation}
g^{(n)} = -(-1)^n 2\left(\frac{\alpha}{2\pi}\right)^{n} \frac{m^2}{s}
U^{(n)}_g ~~,
\label{gn}
\end{equation}
with

\begin{equation}
U^{(n)}_g = \int_{\lambda}^1 d\beta 2\beta(1 - \beta)
F^{(n)}(\beta/\lambda) .
\label{ug}
\end{equation}

In the DLA,

\begin{equation}
\int_{\lambda}^1 d\beta 2\beta(1 - \beta) \ln^n(\beta/\lambda) =
\ln^n(1/\lambda) .
\label{intl}
\end{equation}

Therefore

\begin{equation}
U^{(n)}_g \approx F^{(n)}(1/ \lambda) .
\label{ugf}
\end{equation}

On the other hand, Eq.~(\ref{uf}) reads that

\begin{equation}
\frac{\partial U^{(n)}_f}{\partial \ln(1/\lambda)} =
F^{(n)}(\ln(1/\lambda)) .
\label{fr}
\end{equation}

Combining Eqs.~(\ref{ugf}) and (\ref{fr}), we conclude that in order
$\alpha^n$

\begin{equation}
U^{(n)}_g = \frac{\partial U^{(n)}_f}{\partial \ln(1/\lambda)} .
\label{uggn}
\end{equation}

Summing up (\ref{ugg}) over $n$ leads to
the following obvious relation between $U_f$ and $U_g$,
with all orders in $\alpha$ accounted for  :

\begin{equation}
U_g = \frac{\partial U_f}{\partial \ln(1/\lambda)},
\label{ugg}
\end{equation}
with $U_f = f$ and the Sudakov form factor $f$
is given by
Eq.(\ref{3}).\\
When we put again $\mu = m$,

\begin{equation}
g = g^{(1)}(\rho) U_f(\rho) =
 -2\frac{1}{\rho} \frac{\partial U_f}{ \partial \ln(\rho) }
= -2\frac{\partial f}{\partial \rho}
\label{fg}
\end{equation}
where $\rho = 1/\lambda = s/m^2$.

Therefore, the double-logarithmic asymptotics of Eq.(\ref{1})
for the interaction of electron
with the external electromagnetic field in the kinematical region
(\ref{2}) is~\footnote{The exponentiation of the radiative corrections 
to $g^{(1)}$ in DLA based on was suggested first 
in work \cite{br} where $g$ was calculated in two loops.}

\begin{equation}
\Gamma_{\mu} = \left( \gamma_{\mu} +
\frac{\sigma_{\mu \nu} q_{\nu}}{m} \frac{\partial}{\partial \rho}\right)
\exp\left(-\frac{\alpha}{4\pi} \ln^2\rho\right)
\label{gamma}
\end{equation}

We have obtained the relation (\ref{fg}) between form factors
$f$ and $g$ by direct resummation of the leading contributions of
Feynman graphs to all orders in $\alpha$. However, Eq.~(\ref{gamma})
can be also obtained in much easier way, using the infrared evolution
equation (IREE) method proposed in \cite{efl} for calculating
QCD form factors in DLA. We use this method in Sect.~3 for
obtaining form factor $g$ for quark. Below we demonstrate how it works
in QED,
obtaining again form factor $g$  for electron.  \\
First let us notice that we dropped the electron mass $m$ in
 denominators of integrals
of
Eqs.~(\ref{mn})-(\ref{master}) and  replaced it
by IR cut-off $\mu$ just for the sake of simplicity of these
expressions. It is a consistent procedure in DLA.
However,  it was
not necessary. When both $m$ and $\mu$ are kept ,
\begin{equation}
\label{fgmmu}
g = g^{(1)}(s/\mu^2)U_f(s/\mu^2)
\end{equation}
instead of Eq.~(\ref{fg}). Eqs.~(\ref{fg}) and (\ref{fgmmu}) are identical with
DL accuracy.
However Eq.~(\ref{fgmmu}) reads that
the form factor $g$ of the
electron can be presented as the product of the
single-logarithmic first-loop form factor $g^{(1)}$ (which does not
depend on $\mu$)
 and the $\mu$- dependent
 function $U_f$ presenting DL contributions of the
 other, ``soft'' photons. The factorization
\begin{equation}
\label{anzatse}
g = g^{(1)}(s/m^2)U(s/\mu^2)~~,
\end{equation}
with an unknown function $U(s/\mu^2)$
could be deduced from lower-loop  calculations. After that
 one must
 calculate $U$ in DLA, which can be done
easily using the  IREE  method. Indeed,
although there is no ordering between transverse momenta of
virtual photons in DLA,  the whole phase space
of transverse momenta $k_{i\perp}$ of virtual photons $(i = 1,....)$
can be regarded as a superposition of the regions, each of them
corresponds to a specific ordering between $k_{i\perp}$.
Therefore there is always
the photon $i_m$ with minimal $k_{i_m\perp} \equiv k_{\perp}$ (which we
call the softest photon) in each of those regions.
For such softest photon, the lowest limit
of integrating is $\mu$ whereas limits for integrating over other
$k_{i\perp}$  do not depend on $\mu$. According to
the results of \cite{efl}
the DL contribution of the softest photon  can be
factorized in a certain sense. Namely, $k_{\perp}$ acts as a new
IR cut-off for integrating over other
photon momenta whereas integrating over the
longitudinal Sudakov parameters of the softest photon
can be done like in the first-loop
case, without involving other photon momenta. Therefore the IREE
for $g$ is

\begin{equation}
\label{ireeqed}
g^{(1)}(s/m^2)U(s/\mu^2) =  g^{(1)}(s/m^2) -
g^{(1)}(s/m^2) \left[ \left(
\frac{\alpha}{2\pi}\right)
\int_{\mu^2}^s
\frac{dk^2_{\perp}}{k^2_{\perp}}\ln(s/k^2_{\perp}) \right]
U(s/k^2_{\perp})
\end{equation}
where the first term in rhs is the first loop contribution and 
the expression in the square brackets corresponds to the one-loop
integration over the softest virtual photon momentum. 
Differentiating Eq.~(\ref{ireeqed}) over $\mu$ leads immediately
to

\begin{equation}
\label{ireeu}
 \frac{\partial U}{\partial\ln (s/\mu^2)} = -\frac{\alpha}{2\pi}
\ln(s/\mu^2) U
\end{equation}
with obvious solution $U = exp\left( (-\alpha/4\pi)\ln^2(s/\mu^2) \right)$
which coincides with Eq.~(\ref{fg}).

\section{Elastic form factors of the quark}

In the previous section we have shown that
the form factor $g$ of the
electron can be presented as the product of the
single-logarithmic first-loop contribution of a``hard'' photon
and the independent DL
contributions from the other ``soft'' photons.
This factorization property leads to the exponentiation of the one-loop DL
contribution.
We use below this picture for generalizing (\ref{gamma}) to QCD.

The main difference between QCD and QED is that the quark-gluon interaction
has an additional colour matrix factor. For the one-loop contribution of
Fig.~\ref{oneloop} this leads to an additional colour factor $C_F$
%($C_F=N/2-1/2N, N=3$)
in Eqs.(\ref{f1}),(\ref{a}).
In the next orders the colour factors of different graphs in
Fig.~\ref{mloop} are different: for example  in order $\alpha^2_s$
the colour factors are
$C_F^2$ for the ladder graph and $C_F(-1/2N)=C_F^2-C_F(N/2)$ for
the non-ladder one. In the sum, the contributions proportional to $C_F^2$
combine to reproduce the same result as in QED, with just the replacement
of the electromagnetic charge $\alpha$ of the electron by the colour
charge $\alpha_s C_F$ of the quark. The remaining DL contributions
which are proportional to $C_F(N/2)$ are cancelled by the QCD graphs
with three-gluon vertices.
The generalization of this property to many-loop DL contributions for the
case of the form factor $f_q$ was done in \cite{sudqcd}:

\begin{equation}
f_q = \exp\left[-\frac{\alpha_s C_F}{4\pi} \ln^2\rho~ \right] .
\label{fq}
\end{equation}

\begin{figure}
\begin{center}
\begin{picture}(420,160)
\put(0,10){
\epsfbox{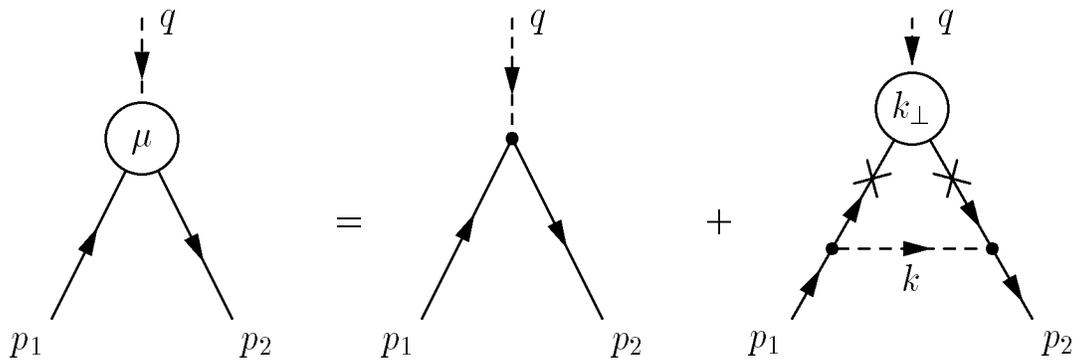}
}
\end{picture}
\end{center}
\caption{The IREE equation for the elastic form factor $f$ of electron and quark.}
\label{fqiree}
\end{figure}

The easiest way to prove the exponentiation of DL radiative corrections
for the quark form factors $f_q$ and $g_q$ is to exploit the
QCD generalization \cite{efl}  of the Gribov bremsstrahlung theorem \cite{g}
and to
use the method of constructing the IREE~\cite{efl}. According to it,
the DL contribution of the virtual photon/gluon with the
smallest $k_{\perp}$ can be separated out of the total amplitude as an
independent factor and the amplitude can be factorized into the product
of this factor and the same amplitude with the new infrared cutoff
$k_{\perp}$ instead of $\mu$, see Fig.~\ref{fqiree}.
For $f_q$ this approach immediately results in the following IREE:

\begin{equation}
\label{irf}
f_q\left(\frac{-q^2}{\mu^2}\right) = 1 - \frac{\alpha_s C_F}{2\pi}
\int_{\mu^2}^{-q^2} \frac{dk^2_{\perp}}{k^2_{\perp}}
\ln\left(\frac{-q^2}{k^2_{\perp}}\right) f_q\left(\frac{-q^2}{k^2_{\perp}}\right)
\end{equation}
where the unity stands for the Born term and $\ln(-q^2/k^2_{\perp})$
in the integrand comes from integration over the longitudinal
components of $k$.  Eq.~(\ref{fq}) is just the solution to 
Eq.~(\ref{irf}).

Let us construct the IREE for the form factor $g_q$. Motivated by
calculations in the lowest orders in $\alpha_s$, we assume that

\begin{equation}
\label{anz}
g_q = g_q^{(1)}\left(\frac{-q^2}{m^2}\right)
U\left(\frac{-q^2}{\mu^2}\right)
\end{equation}
where $g_q^{(1)} = -(\alpha_sC_F/\pi)\ln(-q^2/m^2)$ is the one-loop
value for $g_q$ and $U$ is an unknown DL
function. It is essential that only $U$, not $g_q^{(1)}$, depends on
the infrared cut-off (recall that we put $\mu = m$ only in the
final expressions). Now applying the arguments of
ref.~\cite{efl}, reproduced above, in order to construct the IREE for $g_q$,
we obtain (cf. Eq.~(\ref{ireeqed}))

\begin{equation}
\label{irg}
g_q^{(1)}\left(\frac{-q^2}{m^2}\right) U\left(\frac{-q^2}{\mu^2}\right) =
g_q^{(1)}\left(\frac{-q^2}{m^2}\right) -\frac{\alpha_s C_F}{2\pi}
\int_{\mu^2}^{-q^2} \frac{dk^2_{\perp}}{k^2_{\perp}}
\ln\left(\frac{-q^2}{k^2_{\perp}}\right)
g_q\left(\frac{-q^2}{m^2}\right)U\left(\frac{-q^2}{k_{\perp}^2}\right)~~.
\end{equation}

Comparing Eqs. (\ref{irf}) and (\ref{irg}), we conclude that

\begin{equation}
\label{fgqcd}
g_q = - 2 \frac{\partial f_q}{\partial\rho}
\end{equation}
just as in QED.

In other words, the whole effect of replacement of the electromagnetic
gauge group $U(1)$ by $SU(3)$ is the replacement of
$\alpha$ in Eq.~(\ref{fg})  by $\alpha_s C_F$ and
we arrive at the following expression for
the vertex $\Gamma^{(q)}_{\mu}$ of the quark in an external
electromagnetic field in the kinematical region (\ref{2}):

\begin{equation}
\Gamma^{(q)}_{\mu} = \left[ \gamma_{\mu} +
\frac{\sigma_{\mu \nu} q_{\nu}}{m} \frac{\partial}{\partial \rho}\right]
\exp\left[-\frac{\alpha_s C_F}{4\pi} \ln^2\rho~\right] ~.
\label{quark}
\end{equation}

\section{Inelastic form factors}

The exponential fall of $f$ and $g$ means that non-radiative (elastic)
scattering of an electron or quark is suppressed and therefore  radiative
(inelastic) scattering prevails at large $-q^2$. Indeed, in the
kinematical region (\ref{region}) bremsstrahlung processes are known to
become essential both in QED and in QCD. In the present section we show
that accounting for both gluon and photon bremsstrahlung accounted for
in DLA does
not change the relation between the form factors we obtained in the
previous sections. In the first place let us discuss the case when
electron scattering
by an external electromagnetic field, considered in section 2, is
accompanied by emission of bremsstrahlung photons with
momenta $k_i$ (i=1,..,n). At sufficiently small $k_i$, the scattering
amplitude for the process is

\begin{equation}
\label{inelqed}
\Gamma^{(n)}_{\mu} = B_1B_2...B_n \left[\gamma_{\mu}F_{(n)} -
\frac{\sigma_{\mu \nu}q_{\nu}}{2m}G_{(n)}\right]
\end{equation}
where the photon bremsstrahlung factors $B_i$ are given by Eq. (\ref{5}),
and $F_{(n)}$, $G_{(n)}$ are the inelastic form factors of the electron.
Besides $q^2$, these form factors may also depend on the photon
momenta $k_i$.
Again we use the method of ref.~\cite{efl} to obtain a relation between
$F_{(n)}$ and $G_{(n)}$ and therefore we have to introduce the infrared
cutoff $\mu$ as a minimal value for the transverse momenta of all the
photons involved,
both virtual and emitted. After that we must use again the
generalization
of the  Gribov bremsstrahlung theorem see \cite{efl} like it was done
in previous Sections.

\begin{figure}
\begin{center}
\begin{picture}(450,160)
\put(0,10){
\epsfbox{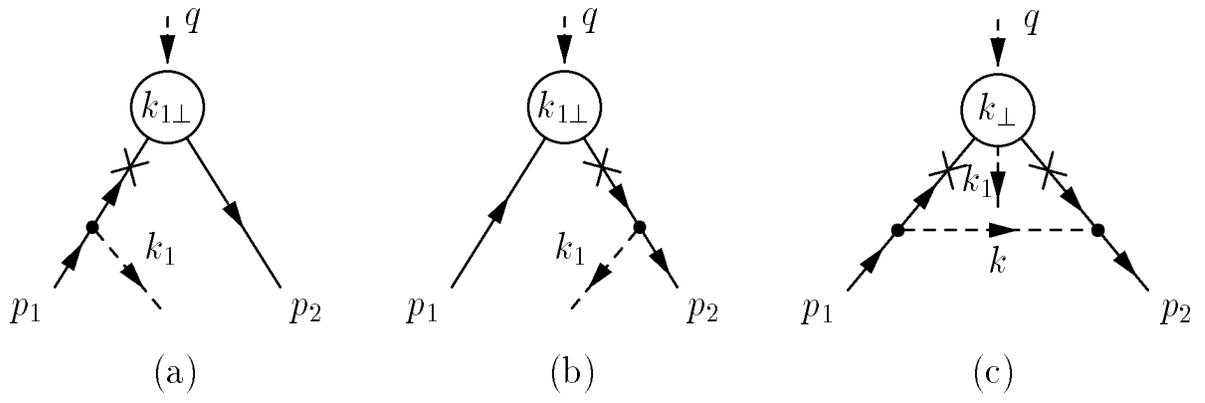}
}
\end{picture}
\end{center}
\caption{Graphs for inelastic form factors of electron.}
\label{bremss}
\end{figure}

For the sake of simplicity, let us begin with $n=1$ in Eq.~
(\ref{inelqed}), considering the emission of a single
bremsstrahlung photon.  In this case one can represent the scattering
amplitude $\Gamma_{\mu}^{(n)}$ by the graphs in Fig.~\ref{bremss}.
As in Fig.~\ref{fqiree} the blob means that radiative
corrections to all orders in $\alpha$ are taken into account in the
DLA.  As the transverse momenta  of the virtual photons are
ordered in the DLA, let us denote by $k_{\perp}$ the smallest of them. We have
to integrate over $k_{\perp}$. In the DLA the region of integration over
$k_{\perp}^2$ runs from $\mu^2$ to $\sim -q^2$. Let us divide it into
two subregions:  the region where $k_{\perp}$ is the smallest

\begin{equation}
\label{k}
k_{\perp} < k_{1\perp}
\end{equation}
and the region where $k_{1\perp}$ is the smallest

\begin{equation}
\label{k1}
k_{\perp} > k_{1\perp} .
\end{equation}

According to the results of \cite{efl} the graphs in
Fig.~\ref{bremss}a,~b correspond to $F_{(1)}$ in the region (\ref{k1})
whereas the
graph in Fig.~\ref{bremss}c corresponds to $F_{(1)}$ in the region (\ref{k}).
Being on mass-shell (the propagators marked by crosses), all the blobs in
Fig.~\ref{bremss} are gauge
invariant. The blobs in Fig.~\ref{bremss}a,~b have infrared cutoff
$k_{1\perp}$ and the blob in Fig.~\ref{bremss}c has infrared cutoff
$k_{\perp}$. Dropping the common factor $B_1$ we obtain the following
IREE for $F_{(1)}$:

\begin{equation}
\label{eqF}
F_{(1)}(q,k_1,\mu) = f\left(\frac{-q^2}{k_{1\perp}^2}\right) -
\frac{\alpha}{2\pi} \int_{\mu^2}^{k_{1\perp}^2}
\frac{dk^2_{\perp}}{k^2_{\perp}} \ln\left(\frac{-q^2}{k^2_{\perp}}\right)
F_{(1)}(q,k_1,k_{\perp})  .
\end{equation}

Differentiating Eq.~(\ref{eqF}) with respect to $\mu$
immediately yields that the
solution does not depend on
$k_1$ and coincides with the elastic form factor $f(-q^2/\mu^2)$,
Eq.~(\ref{3}). This result was obtained earlier in \cite{gor} by
direct calculation of Feynman graphs in DLA.

As the soft bremsstrahlung of the photon $k_1$ does not influence the
spinor structure of the blobs in Fig.~\ref{bremss}, we can easily
generalize the IREE (\ref{eqF}) to the case of the form factor $G_{(1)}$,
just substituting it for $F_{(1)}$ and $g$ for $f$. Due to the infrared
stability of the one-loop result $g^{(1)}$ one can consider it to be a
common factor independent of $\mu$.

Using
%the relation

%\begin{equation}
%\label{gg1f}
%g\left(\frac{-q^2}{\mu^2}\right) = g^{(1)}\left(\frac{-q^2}
%{k_{1\perp}^2}\right)
%f\left(\frac{-q^2}{\mu^2}\right)
%\end{equation}
 the ansatz

\begin{equation}
\label{u1}
 G_{(1)} = g^{(1)}(s/m^2)U_{(1)}(s/\mu^2) ,
\end{equation}
we immediately obtain $U_{(1)} = F_{(1)}$.  Repeating the same method
for $n = 2,3,..$ emitted photons, we conclude that

\begin{equation}
\label{FG}
G_{(n)} = -2\frac{\partial F_{(n)}}{\partial\rho} .
\end{equation}

The fact that neither $F_{(n)}$ nor $G_{(n)}$ depends on the
bremsstrahlung photon momenta implies factorization of the
bremsstrahlung and a Poisson energy spectrum for the soft
photons.

Now let us turn to QCD and consider the process where instead of the
electron a quark interacts with an external electromagnetic field, and
take into account emission of $n$ bremsstrahlung gluons.  As
each emitted gluon may emit other gluons, producing gluon cascades, the
gluon bremsstrahlung factors and the whole picture are more
complicated.
As was shown in \cite{ef}, in DLA all the bremsstrahlung gluons inside
a cascade, or jet, are strongly ordered in energies and emission
angles, relative to their parent gluons. This is a result of
destructive interference or the coherence effect. Therefore one has
to use different representations for the scattering amplitude
$\tilde{\Gamma}_{\mu}^{(n)}$ for each DL kinematical region, accounting
for the various ways to group $n$ gluons into jets. In each such
region one can expect that the scattering amplitude
$\tilde{\Gamma}_{\mu}^{(n)}$ takes the simple form of a product of
the gluon bremsstrahlung factors $\tilde{B_i}$ and the inelastic form
factors $\tilde{F}_{(n)}$ and $\tilde{G}_{(n)}$

\begin{equation}
\label{inelqcd}
\tilde{\Gamma}_{\mu}^{(n)} = \tilde{B}_1...\tilde{B}_n
\left[\gamma_{\mu} \tilde{F}_{(n)} -
\frac{\sigma_{\mu\nu}q_{\nu}}{2m} \tilde{G}_{(n)}\right] .
\end{equation}

\begin{figure}
\begin{center}
\begin{picture}(360,160)
\put(0,10){
\epsfbox{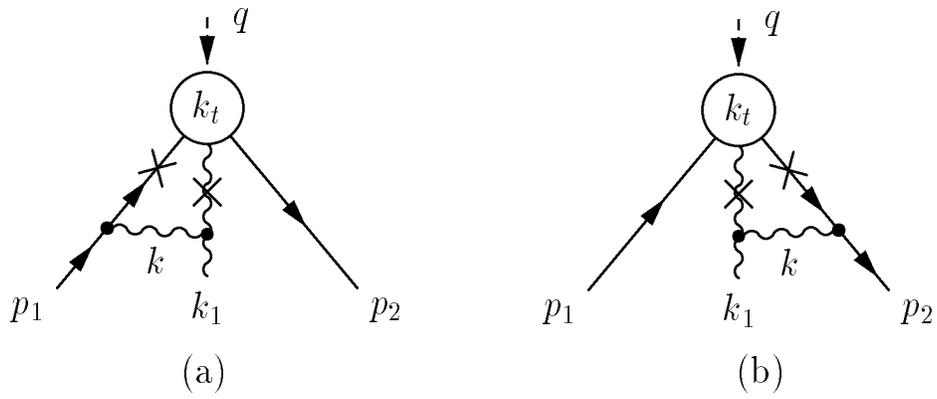}
}
\end{picture}
\end{center}
\caption{3-gluon graphs for inelastic form factors of quark.}
\label{ggg}
\end{figure}

To make the problem clearer let us consider the simplest case $n=1$.
Then besides the graphs of Fig.~\ref{bremss} the two new graphs of
Fig.~\ref{ggg} give DL contributions to $\tilde{F}_{(1)}$ and the IREE
(\ref{eqF}) takes the following form:

\begin{equation} \label{irf1}
\tilde{F}_{(1)}(q,k_1;\mu) =
f\left(\frac{-q^2}{k_{1\perp}^2}\right) -\frac{\alpha_s}{2\pi}
\int_{\mu^2}^{k^2_{1\perp}} \frac{dk^2_t}{k^2_t} \left[C_F
\ln\left(\frac{-q^2}{k^2_t}\right) + \frac{N}{2}
\ln\left(\frac{2(p_1k_1)}{k^2_t}\right) + \frac{N}{2}
\ln\left(\frac{2(p_2k_1)}{k^2_t}\right) \right]
\tilde{F}_{(1)}(q,k_1;k_t)
\end{equation}
where we have accounted for the different colour factors for the graphs
in Fig.~\ref{bremss}c and in Fig.~\ref{ggg}. It is worthwhile to mention that
in each term of the integrand in Eq.~(\ref{irf1}) $k_t$ must be defined
separately, with respect to the momenta $p_1$ and $p_2$ for the first term,
$p_1$ and $k_1$ for the second term and $p_2$, $k_1$ for the last term
(see \cite{efl} for details).

Due to the kinematical relation

$$2(p_1k_1) 2(p_2k_1) = (-q^2) k^2_{1\perp}$$
we obtain the following solution to Eq.~(\ref{irf1})

\begin{equation} \label{ff1}
\tilde{F}_{(1)} = \exp\left( -\frac{\alpha_s}{4\pi}
\left[ C_F\ln^2\left(\frac{-q^2}{\mu^2}\right) +
\frac{N}{2}\ln^2\left(\frac{k^2_{1\perp}}{\mu^2}\right) \right]\right) .
\end{equation}

In contrast to the inelastic form factor of the electron, Eq.(\ref{3}),
the form factor of the quark $\tilde{F}_{(1)}$ depends on $k_1$. This implies
the violation of both factorization and the Poisson energy spectrum
for gluon bremsstrahlung.

The expression obtained for $\tilde{F}_{(1)}$ is the exponentiation of the
one-loop contribution. The same exponentiation was shown in
\cite{efl} to be true for the general case of $n$ bremsstrahlung gluons,
$\tilde{F}_{(n)}$.

Obviously, if we keep the ansatz

\begin{equation}
\label{gffn}
G_{(n)} = g^{(1)}\left(q,k_1,m\right)
\tilde{U}_{(n)}(q,k_1,\mu) ,
\end{equation}
the above argument holds for constructing the IREE of
the DL function $\tilde{U}_{(n)}(q,k_1,...,k_n;\mu)$ and for it being equal
to $\tilde{F}_{(n)}(q,k_1,...,k_n;\mu)$, which results in the same relation,
Eq.~(\ref{FG}).

\section{Conclusion}

We have calculated the electromagnetic form factors of the
electron and quark in the asymptotic regime where the momentum $q$
transferred to the electron or quark is much greater than their masses.
The Eqs.~(\ref{gamma}) and (\ref{quark}) that
we have obtained predict  an exponential fall for the form factors
when $q^2$ increases. This corresponds
to the suppression of non-radiative scattering, without
photon bremsstrahlung,  at high energies.
When photon bremsstrahlung is taken into account in the DLA,
in the expressions for the radiative (inelastic) form factors
Eqs.~(\ref{gamma}) and (\ref{quark})
are multiplied  by the bremsstrahlung
factors $B_i$ defined in Eq. (\ref{5}).
All the form factors in Eqs. (\ref{gamma}) and (\ref{quark})
are independent of momenta of the emitted photons.
This means that taking into account the form factor $g$
does not violate the Poisson energy spectrum for the photon
bremsstrahlung in the
double logarithmic approximation. In contrast, the inelastic
form factor that accounts for gluon bremsstrahlung depends, in the DLA,
both on the
emitted gluon momenta and on the structure of each of the gluon cascades
\cite{ef,efl}. This makes it possible to obtain simple expressions for
the inelastic form factor in QCD only separately for every kinematical
region corresponding to a certain ordering of the emission energies and
angles \cite{ef}. Still, in the DLA, even with that complication, the
relation Eq.~(\ref{FG}) holds also for the two quark form factors.

\section{acknowledgements}

We are grateful to G. Altarelli, M. Mangano, N. Nikolaev and D. Soper for
useful discussions. The work was supported in part by the grant
INTAS-97-30494.

%%%%%%%%%%%%%%%%%%%%%%%%%%%%%%%%
%%%%%%%%%%%%%%%%%%%%%%%%%%%%%%%%
%%%%%%%%%%%%%%%%%%%%%%%%%%%%%%%%
%%%%%%%%%%%%%%%%%%%%%%%%%%%%%%%%
%%%%%%%%%%%%%%%%%%%%%%%%%%%%%%%%

\end{document}